\documentclass[showpacs,aps,twocolumn,prl]{revtex4}
\usepackage{graphicx}
\begin{document}
\bibliographystyle{apsrev}

\title{Rank distributions of words in additive many-step Markov
chains and the Zipf law}
\author{K. E. Kechedzhy}
\affiliation{Department of Physics, Kharkov National University, 4
Svoboda Sq., Kharkov 61077, Ukraine}

\author{O. V. Usatenko, V. A. Yampol'skii
\footnote[1]{yam@ire.kharkov.ua} }
\affiliation{A. Ya. Usikov Institute for Radiophysics and Electronics \\
Ukrainian Academy of Science, 12 Proskura Street, 61085 Kharkov,
Ukraine}

\begin{abstract}
The binary many-step Markov chain with the step-like memory
function is considered as a model for the analysis of rank
distributions of words in stochastic symbolic  dynamical systems.
We prove that the envelope curve for this distribution obeys the
power law with the exponent of the order of unity in the case of
rather strong persistent correlations. The Zipf law is shown to be
valid for the rank distribution of words with lengths about and
shorter than the correlation length in the Markov sequence. A
self-similarity in the rank distribution with respect to the
decimation procedure is observed.

\end{abstract}

\maketitle

The rank distributions (RD) in the stochastic systems attract the
attention of specialists in the physics and many other fields of
science because of their universal power-law character (the
so-called Zipf law (ZL)~\cite{zipf}. Discovered originally for the
RD of words in the natural languages, the ZL was later observed in
the rank distributions of other objects, such as distributions of
"words" in the DNA sequences~\cite{dna}, PC codes~\cite{dna},
capitals of stock market players~\cite{cald} (in economics, the
Zipf law in slightly different form is known as Pareto's principle
or the 80-20 rule~\cite{paret}), the population of cities, areas
occupied by countries, masses of cosmic objects etc
(see~\cite{trub}). In spite of a lot of endeavors to describe this
phenomenon analytically, a deep insight into the problem has not
so far been gained.

To define the rank distribution of some objects in a definite
sequence, it is necessary to establish a correspondence between
objects and their frequencies of appearing in the sequence and to
arrange the objects in ascending order of these frequencies. A
choice of the model for analytical description of the Zipf law in
RD is rather ambiguous because of diversity of the systems where
it occurs~\cite{trub}. Here the principal question arises
concerning the way of defining the objects that are involved in a
competition according to the frequency of their occurring in the
sequence under consideration. There exist two principally
different approaches to this problem. The first of them consists
in considering the objects as a priori equivalent, i.e. having
"the same rights" in the competition. The Zipf-law in rank
distributions in such models appears only due to the correlations
that are present in the sequence. The rank distribution of the
triplets in the DNA sequences can serve as a vivid example of the
real systems for which this approach is essential
(see~\cite{dna}).

The second approach deals with sequences where the correlations
does not play an essential role. However, the competitive objects
have a priori nonequal chances to take a given place in the
sequence. Mandelbrot`s models~\cite{mand1,mand2}, the Kanter and
Kessler model~\cite{kk} and other models are constructed on the
basis of the choice of the a priori nonequivalent competitors, and
specifically this non-equivalency is a reason for obeying the rank
distributions to the Zipf law. For example, the non-equivalency of
the words in literary texts is caused by their different lengths
and, consequently, by their different statistical weight that is
defined by a number of characters in the word and the capacity of
alphabet.

The nature of the real objects that satisfy the Zipf law does not
furnish sufficient arguments for giving preference to one of the
discussed approaches. It is possible that both approaches describe
different mechanisms of forming the power-law rank distributions.

In the present paper, we suggest an analytically solvable model of
many-step Markov chain where the rank distribution of different
$L$-words ($L$ consequent symbols in the chain) of a definite
length $L$ is examined. Since the words are equal in their length,
the rank distribution in this system occurs as a result of the
correlations. In other words, our model is based specifically on
the first approach to the choice of the competitors in the
sequence. The study that has been carried out allowed us to reveal
the relation of the rank distributions to the existent
correlations in the system. The speculations about the connection
of the Zipf law to the \textit{long-range correlations} were
expressed clearly by a number of authors~\cite{den,czir}. We have
demonstrated that the \textit{short-range correlations} can also
provide the appearance of the Zipf law.

We have analytically studied the rank distributions of words of
certain length $L$ in the Markov chains. If a Markov chain
possesses the one-step-like memory function considered in
Ref.~\cite{uya}, this distribution is shown to be of the
many-step-like form. In the case of strong correlations, the
envelope curve for the rank distribution obeys the power-law
behavior with the exponent of the order of unity, i.e. the
distribution is described by the Zipf law. The obtained results
provide us with a sufficient amount of information to clarify the
origin of Zipf's law. In particular, we have made sure that the
correlations of symbols within the competitive words is sufficient
for the appearance of the Zipf law in their rank distributions.

The suggested approach to the problem of the Zipf law is expedient
because we are provided by the theoretical parameters that affect
both the character of correlations and the rank distribution of
words occurring in the Markov chain. Due to this circumstance, we
could examine the relationship between the rank distributions and
the correlation properties of the system.

Let us consider a homogeneous stationary unbiased binary sequence
of symbols, $a_{i}=\{0,1\}$, and define the word as a set of
sequential symbols of definite length $L$. Different words are
obtained by progressively shifting a window of the length $L$ by
one symbol in the sequence. The rank distribution of words is a
relationship connecting the probability $W$ of certain word
occurring to the corresponding rank. The words are ordered in
ascending rank order, $W(1) \geq W(2) \dots \geq W(2^L)$. Our
sequence is the $N$-step Markov chain with the step-like memory
function. This means that the conditional probability $P(a_{i}\mid
a_{i-N},a_{i-N+1},\dots ,a_{i-1})$ of definite symbol $a_i$
occurring (for example, $a_i =0$) after symbols
$a_{i-N},a_{i-N+1},\dots ,a_{i-1}$ in the chain is determined by
the equation,
\[
P(a_{i}=0\mid a_{i-N},a_{i-N+1},\dots ,a_{i-1})
\]
\begin{equation}
=1/2+\mu (1-2k/N). \label{2}
\end{equation}
Here $k$ denotes the number of unities among $N$ symbols,
$a_{i-1}, a_{i-2}, \dots a_{i-N}$, preceding the generated one,
$a_i$, and $\mu$ is the strength of correlations in the sequence,
$-1/2 < \mu < 1/2$. The case with $\mu=0$ corresponds to the
non-correlated random sequence of symbols. The positive (negative)
values of $\mu$ correspond to the persistent (anti-persistent)
correlation (the attraction (repulsion) of symbols of the same
kind).

As was shown in Refs.~\cite{uya,uyakm}, the probability $W$ of
certain $L$-word occurring depends on the number $k$ of unities in
the word with $L \leq N$ but is independent of their arrangement.
It is described by the formula,
\begin{equation}\label{4}
W(k)=W(0)\frac{\Gamma(n+k) \Gamma (n+L-k)}{\Gamma (n) \Gamma
(n+L)},
\end{equation}
with
\[
W(0)=\frac{4^n }{2\sqrt{\pi}}\frac{\Gamma(1/2+n)\Gamma
(n+L)}{\Gamma (2n+L)},
\]
\begin{equation}\label{5}
n=\frac{N(1-2\mu)}{4\mu}.
\end{equation}
Since the probability $W(k)$ does not depend on the arrangement of
symbols within the $L$-word, the specific degeneration takes place
in the Markov chain under study. Another kind of degeneration
arises from the non-bias property of the sequence: the probability
$W(k)$ is symmetric with respect to the change $k \rightarrow
(L-k)$, $W(k)=W(L-k)$. Thus, $2\text{C}_L^k = 2L!/k!(L-k)!$
different words occur with the same frequency $W(k)$. This results
in the step-like form of the rank distribution of the $L$-words
with $L \leq N$. Each step can be labelled by the number $k \leq
L/2$ of unities (or zeros) within them and is characterized by the
length equal to the degeneracy multiplicity $2\text{C}_L^k$. The
right edge of the $k$th step corresponds to the rank $R(k)$ which
is described by the equation,
\begin{equation}\label{6}
R(k)=2\sum\limits_{i=0}^{k}\text{C}_{L}^{i}.
\end{equation}
Indeed, performing the ranking procedure (all words containing
equal numbers of unities $k$ have neighboring ranks) we obtain
this formula. A pair of Eqs.~(\ref{4}) and (\ref{6}) being
considered as a parametrically defined function $W(R)$ represents
the envelope curve passing through the right edges of the steps in
the rank-distribution.

Using the Stirling formula for the Gamma-functions (which is valid
at $L, k, (L-k) \gg 1$) and changing the summation operation in
Eq.~(\ref{6}) by integration, one can easily obtain the asymptotic
expression for the dependence $R(W)$,
\begin{equation}\label{7}
R=2^L \sqrt{\frac{\zeta}{\pi}}
\left(\frac{W}{B}\right)^{-\frac{1}{\zeta}}
\ln^{-1/2}\left(\frac{W}{B}\right)
\end{equation}
with
\[
\zeta=\frac{1}{1 +2n/L},
\qquad B= 4^n \sqrt{\pi}\frac{\Gamma(1/2
+ n)}{\Gamma(n)\Gamma(2n + L)}
\]
\begin{equation}\label{zeta}
\times \left(n + \frac{L}{2} -1 \right)^{2n +L-1} \times \exp(-
2n-L+2).
\end{equation}
The distribution Eq.~(\ref{7}) differs from the usually discussed
power-law form by the logarithmic multiplier only. If one neglects
this weak logarithmic dependence, the Zipf law for the rank
distribution would be obtained from Eq.~(\ref{7}),
\begin{equation}\label{8}
W \propto R^{-\zeta}.
\end{equation}

\begin{figure}[h!]
\begin{centering}
\scalebox{0.85}[0.85]{\includegraphics{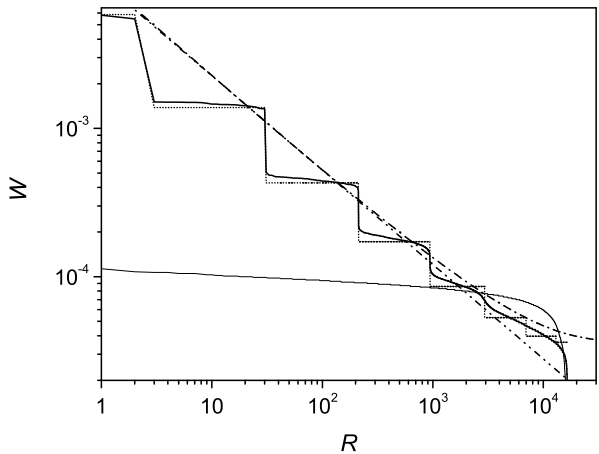}}\caption{The rank
distribution $W(R)$ of the words of the length $L=14$ in the
Markov chain with a step-like memory function, $N=15$, $\mu =
15/46$. The solid line corresponds to the numerical simulations,
dotted line describes the exact distribution obtained from
Eq.~(\ref{4}), dash-dotted line is the plot of envelope
asymptotics Eq.~(\ref{7}), the dash-dot-dotted line describes the
Zipf asymptotics Eq.~(\ref{8}) with $\zeta = 7/11$. Thin solid
line corresponds to anti-persistent correlations obtained at $\mu
= - 15/46$, $N=15$,  $L=14$.}\label{Fig1}
\end{centering}
\end{figure}

The achieved result is demonstrated in Fig.~\ref{Fig1}. The dotted
line shows the plot of the rank distribution obtained from
Eq.~(\ref{4}) at $L=14$, $N=15$, $\mu=15/46$, $n=4$. This plot
passes closely to the solid line, which demonstrates the results
of numerical simulations of the rank distribution of words of the
length $L=14$ in the Markov chain generated with the same
parameters $N$ and $\mu$. The dash-dotted line in this figure is
the envelope curve Eq.~(\ref{7}). \protect

The exponent in Zipf's distribution Eq.~(\ref{8}) is governed by
the ratio $n/L$. In the case of weak correlations, at $n
\rightarrow \infty$, the value of $\zeta$ tends to zero and the
Zipf distribution appears to be destroyed, i.e. all words of the
length $L$ occurs with the almost same probabilities $W=2^{-L}$.
The opposite situation, $n/L \rightarrow 0$, corresponds to the
strong correlations in the Markov chain. Eq.~(\ref{zeta}) shows in
this case that the exponent $\zeta$ tends to unity. The plot of
Zipf distribution Eq.~(\ref{8}) with $\zeta = 7/11$ is
demonstrated by the dash-dot-dotted line in Fig.~\ref{Fig1}.

\begin{figure}[h!!!]
\begin{centering}
\scalebox{0.85}[0.85] {\includegraphics{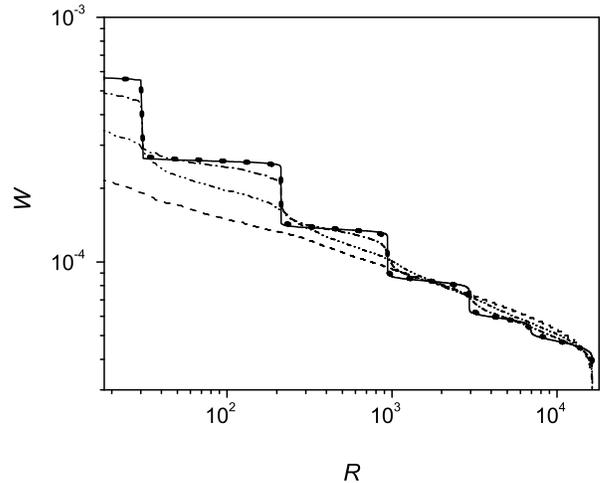}} \caption{ Rank
distributions of the $L$-words with $L=14$ in the $N$-step Markov
sequences reduced by randomly removing some symbols. Solid line
corresponds to the initial sequence possessing $N=32, n=15$.
Symbols corresponding to the decimation parameter $\lambda=2$
($N^\ast=16$) lie almost on the solid one. Other lines correspond
to decimated sequences with $N^\ast <L$. Specifically,
dash-dotted, dash-dot-dotted, and dotted lines correspond to
$N^\ast =8$, $N^\ast =4$, and to $N^\ast =2$,
respectively.}\label{Fig2}
\end{centering}
\end{figure}

According to statements given in the literature, the Zipf law is
associated with the property of scale invariance~\cite{czir}, i.e.
invariance of the slope of the Zipf plot with respect to certain
decimation procedure. An analogous property referred to as the
self-similarity, appears in the frame of the model presented
above. Let us reduce the $N$-step Markov sequence by regularly (or
randomly) removing some symbols and introduce the decimation
parameter $\lambda < 1$ which represents the fraction of symbols
kept in the chain. As is shown in Ref.~\cite{uyakm}, the reduced
chain possesses the same statistical properties as the initial one
but is characterized by the renormalized memory length, $N^{\ast
}$, and the persistence parameter, $\mu ^{\ast }$,
\begin{equation}
N^{\ast }= N\lambda, \qquad \mu ^{\ast }=\mu \frac{\lambda
}{1-2\mu (1-\lambda )}. \label{43}
\end{equation}
Indeed, the conditional probability $p_{k}^{\ast }$ of occurring
the symbol zero after $k$ unities among the preceding $N^{\ast }$
symbols in the reduced chain is described by Eq.~(\ref{2}) where
$N$ and $\mu$ should be replaced by the renormalized parameters
$N^{\ast }$ and $\mu ^{\ast }$. Considering the Zipf law, we are
interested in the invariance of the Zipf plot with respect to the
decimation procedure. According to Eqs.~(\ref{7})
and~(\ref{zeta}), the slope of the Zipf plot depends on the
parameter $n$ only. This parameter does not change after the
transformation $N \rightarrow N^{\ast}$, $\mu \rightarrow
\mu^{\ast}$ (see Eqs.~(\ref{5}) and~(\ref{43})) . As a result, the
Zipf plots for rank distributions of $L$-words with $L<N$,
obtained from the initial and decimated sequences, coincide. This
self-similarity property is demonstrated in Fig.~\ref{Fig2}.
\protect

\begin{figure}[here!]
\begin{centering}
\scalebox{0.85}[0.85] {\includegraphics{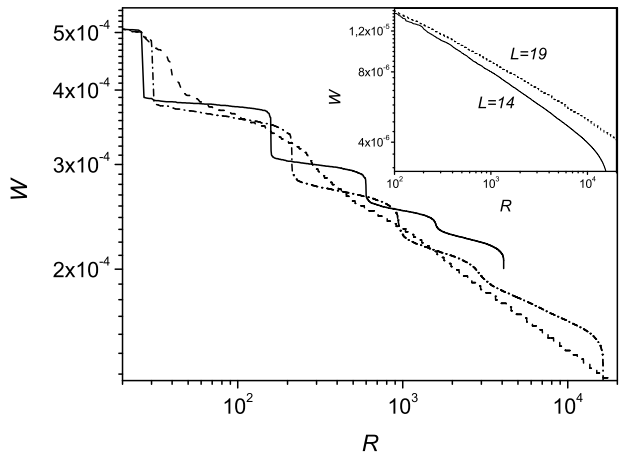}} \caption{The
Zipf plots for $L$-words with $L>N$. Solid, dash-dotted,  and
dashed lines correspond to $L=12, 14, 18$, respectively. The
parameters of the Markov chain are $N=12, \mu = 0.1$. The
phenomenon of step smearing is observed with a growth of the word
length $L$. In the inset: the Zipf plots for $L$-words with $N=4,
\mu=0.1$ at $L>L_{cr}$ (the lengths of words are shown near the
curves).} \label{Fig3}
\end{centering}
\end{figure}

Now, let us study the rank distribution of $L$-words with $L>N$.
This problem is not amenable to analytic calculations, and,
therefore, numerical simulations are applied. In this case, the
above-mentioned degeneration of the probability of word occurring
is non-existent. So, smearing of the steps in the rank
distribution takes place at $L>N$. This smearing occurs gradually
with an increase of the word length $L$ and the steps appear to be
completely smoothed away at high enough values of $L$ (see curves
in Fig.~\ref{Fig3}). This means that the Zipf law describes the
rank distribution itself contrary to its envelope curve at $L<N$.
\protect

It is important to draw attention to the non-monotonous behavior
of the Zipf slope $\zeta$ with an increase in the word length $L$.
As is seen from Eq.~(\ref{zeta}), this parameter increases at
$L<N$. This growth continues at $L>N$ as well but only up to a
certain value of $L=L_{cr}>N$. The maximum of $\zeta$ is observed
at $L=L_{cr}>N$ and then, at $L>L_{cr}$, $\zeta$ starts to
decrease. This decrease is demonstrated in the inset to
Fig.~\ref{Fig3}.

It is necessary to note that the position $L=L_{cr}$ of maximum in
the $\zeta (L)$ dependence is strongly related to the
characteristic correlation length $2l_c + N$ in the Markov chain
being studied. According to Ref.~\cite{uyakm}, the symbols
correlate with each other not only within the memory length $N$
but within the enlarged region $2l_c + N$ where $l_c$ represents
the characteristic attenuation length of the fluctuations. Thus,
the best fitting of the rank distribution of words in the Markov
chain by the power-law curve is achieved if the size of the
competitive words is close to the correlation length.

If the words are shorter than the correlation length, $L<N+2l_c$,
the specific features of the correlations become apparent in the
rank distribution that results in deviations from the Zipf law. In
the system that is considered in this paper, the deviations
manifest themselves in the appearance of the steps in the rank
distribution and in the additional weak logarithmic multiplier in
Eq.~(\ref{7}). Moreover, at very small $N \sim 1$ the rank
distribution deviates significantly from the power law and gets
the exponential shape at $N=1$.

In the opposite limiting case, at $L \gg N+2l_c$, the correlations
over the whole word length disappear and the rank distribution
tends to the constant. It is important to note that specifically
persistent correlations in the Markov chain (that correspond to
the attraction between the same symbols) lead to the pronounced
Zipf law in the rank distribution of words. Indeed, the thin solid
line in Fig.~\ref{Fig1} demonstrates the very weak $W(R)$
dependence for the case of the antipersistent correlations, at
$\mu=-15/46$.

Thus, the obtained results allow us to suggest the following
physical picture of the appearance of the Zipf law. The
correlations should be presented in the system but the noise of
sufficient strength should be imposed on these correlations over
the length of competitive words. Within the considered stochastic
system, this noise is provided by sufficiently strong fluctuations
observed on the scales of the word length. The role of the noise
consists in concealing the specific peculiarities of the
correlations. Owing to the fluctuations, the concrete shape of the
correlations does not appear to be very important. Accordingly,
the Zipf law is a consequence of the rather rapidly damping
persistent correlations of quite arbitrary form, i.e. the global
correlations in the system are not necessary. The Zipf law is a
manifestation of the inner microstructure of the system being a
result of attraction between building blocks of the same kind.

We acknowledge Dr. S. S. Denisov for the helpful discussions.

\end{document}